\documentclass[preprint,showpacs,preprintnumbers,amsmath,amssymb]{revtex4}

\usepackage{bm}
\usepackage{epsfig}

\begin{document}
\begin{widetext}

\noindent{\Large\bf Folding of the Protein Domain hbSBD}
\vskip 5 mm


\noindent{\bf Maksim Kouza$^{1,2}$,
Chi-Fon Chang$^{3}$, Shura Hayryan$^2$, Tsan-hung Yu$^4$,\\ Mai Suan Li$^1$,
Tai-huang Huang$^{4,5}$, and Chin-Kun Hu$^{2,6}$}
\vskip 2 mm

\noindent{\it $^1$Institute of Physics, Polish Academy of Sciences,
Al. Lotnikow 32/46, 02-668 Warszawa, Poland}

\noindent{\it $^2$Institute of Physics, Academia Sinica, Nankang,
Taipei 11529, Taiwan}

\noindent{\it $^3$Genomics Research Center, Academia Sinica, Nankang,
Taipei 11529, Taiwan}

\noindent{\it $^4$Institute of Biomedical Sciences, Academia Sinica, Nankang,
Taipei 11529, Taiwan}

\noindent{\it $^5$Department of Physics, National Taiwan Normal University,
Taipei 11718, Taiwan.}

\noindent{\it $^6$National Center for Theoretical Sciences at
Taipei, Physics Division, National Taiwan University, Taipei
10617, Taiwan}

\end{widetext}


\vskip 5 mm \noindent{\bf ABSTRACT.
The folding of the $\alpha$-helice domain hbSBD of the mammalian
mitochondrial branched-chain $\alpha$-ketoacid dehydrogenase
(BCKD) complex is studied by the circular dichroism technique in
absence of urea. Thermal denaturation is used to evaluate various
thermodynamic parameters defining the equilibrium unfolding, which
is well described by the two-state model with the folding
temperature   $ T_F = 317.8 \pm 1.95$ K and the enthalpy change
$\Delta H_G = 19.67 \pm 2.67$ kcal/mol. The folding is also
studied numerically using the off-lattice coarse-grained Go model
and the Langevin dynamics. The obtained results, including the
population of the native basin, the free energy landscape as a
function of the number of native contacts and the folding
kinetics, also suggest that the hbSBD domain is a two-state
folder. These results are consistent with the biological function
of hbSBD in BCKD.}



\newpage

\noindent
{\Large \bf Introduction}
\vskip 2 mm

\noindent
Understanding the dynamics and mechanism of protein
folding remains one of the  most challenging problems in molecular
biology \cite{Fersht03}.  Single domain $\alpha$ proteins attract
much attention of researchers because most of them fold faster
than $\beta$ and $\alpha\beta$ proteins \cite{Jackson98,Eaton04}
due to relatively simple energy landscapes and one can, therefore,
use them to probe main aspects of the funnel theory
\cite{Wolynes95}. Recently, the study of this class of proteins
becomes even more attractive because the one-state or downhill
folding may occur in some small $\alpha$ proteins
\cite{Munoz02,Munoz04,Ferguson04}.

The mammalian mitochondrial branched-chain $\alpha$-ketoacid
dehydrogenase (BCKD) complex catalyzes the oxidative decarboxylation
of branched-chain $\alpha$-ketoacids derived from leucine, isoleucine
and valine to give rise to branched-chain acyl-CoAs.
In patients with inherited maple syrup urine disease,
the activity of the BCKD complex is deficient, which
is manifested by often fatal acidosis and mental retardation \cite{ccf1}.
 The BCKD multi-enzyme complex (4,000 KDa in size) is organized about
a cubic 24-mer core of dihydrolipoyl transacylase (E2), with multiple
 copies of hetero-tetrameric decarboxylase (E1), a homodimeric
dihydrogenase (E3), a kinase (BCK) and a phosphatase attached
through ionic interactions.  The E2 chain of the human BCKD
complex, similar to other related multi-functional enzymes
\cite{ccf2}, consists of three domains:  The amino-terminal
lipoyl-bearing domain (hbLBD, 1-84), the interim E1/E3
subunit-binding domain (hbSBD, 104-152) and the carboxy-terminal
inner-core domain. The structures of these domains serve as bases
for modeling interactions of the E2 component with other
components of $\alpha$-ketoacid dehydrogenase complexes. The
structure of hbSBD (Fig. 1) has been determined by NMR
spectroscopy, and the main function of the hbSBD is to attach both
E1 and E3 to the E2 core \cite{ccf3}. The two-helix structure of
this domain is reminiscent of the small protein BBL
\cite{Ferguson04} which may be a good candidate for observation of
downhill folding \cite{Munoz02,Munoz04}. So the study of hbSBD is
interesting not only because of the important biological role of
the BCKD complex in human metabolism but also for illuminating
folding mechanisms.

From the biological point of view, hbSBD could be less stable than
hbLBD and  one of our goals is, therefore, to check this by the
circular dichroism (CD) experiments. In this paper we study the
thermal folding-unfolding transition in the hbSBD by the CD
technique in the absence of urea and pH=7.5. Our thermodynamic
data do not show evidence for the downhill folding and they are
well fitted by the two-state model. We obtained folding
temperature $ T_F = 317.8 \pm 1.95$ K and the transition enthalpy
$\Delta H_G = 19.67 \pm 2.67$ kcal/mol. Comparison of such
thermodynamic parameters of hbSBD with those for hbLBD shows that
hbSBD is indeed less stable as required by its biological
function. However, the value of $\Delta H_G$ for hbSBD is still
higher than those of two-state $\alpha$ proteins reported in
\cite{Eaton04}, which indicates that the folding process in the
hbSBD domain is highly cooperative.


From the theoretical point of view it is very interesting to
establish if the two-state foldability of hbSBD can be captured by
some model. The all-atom model would be the best choice for a
detailed description of the system but the study of hbSBD requires
very expensive CPU simulations. Therefore we employed the
off-lattice coarse-grained Go-like model \cite{Go,Clementi00}
which is simple and allows for a thorough characterization of
folding properties. In this model amino acids are represented by
point particles or beads located at positions of $C_{\alpha}$
atoms. The Go model is defined through the experimentally
determined native structure \cite{ccf3}, and it captures essential
aspects of the important role played by the native structure
\cite{Clementi00,Takada99}.

It should be noted that
the Go model by itself can not be employed to ascertain the two-state behavior
of proteins.
However, one can use it in conjunction with experiments providing
the two-state folding because this model does not {\it always} provide
the two-state behavior as have been clearly shown in the seminal work
of Clementi {\it et al.} \cite{Clementi00}. In fact,
the Go model correctly captures not only the two-state folding of
proteins CI2 and SH3 (more two-state Go folders may be found
in Ref. \cite{Koga01})
but also intermediates of the three-state folder
barnase, RNAse H and CheY \cite{Clementi00}.
The reason for this is that
the simple Go model ignores the energetic frustration but it still takes
the topological frustration into account.
Therefore, it can capture intermediates 
that occur due to topological constraints but not those
emerging from the frustration of the contact interactions. 

 With the help of Langevin dynamics
simulations and the histogram method \cite{Ferrenberg89} we have
shown that, in agreement with our CD data, hbSBD is a two-state
folder with a well-defined transition state (TS) in the free
energy landscape. The two helix regions were found to be
highly structured in the TS. The two-state behavior of hbSBD  
is also supported by our kinetics study
showing that the folding kinetics follows the single exponential scenario.
The two-state folding obtained in our simulations suggests that for hbSBD
the topological frustration is more important than the energetic factor.

 The dimensionless quantity, $\Omega _c$
\cite{KlimThirum98FD}, which characterizes the structural
cooperativity of the thermal denaturation transition was computed
and the reasonable agreement between the CD experiments and Go
simulations was obtained. Incorporation of side chains may give a
better agreement \cite{KlimThirum98FD,Li_Physica05} but this
problem is beyond the scope of the present paper.

\vskip 6 mm
\noindent{\Large \bf {Materials and Methods}} \vskip 2 mm

\noindent{\bf Sample Preparation}

\vskip 2 mm

\noindent
 hbSBD protein was purified from the BL21(DE3) strain of
\textit{E. coli }containing a plasmid that carried the gene of
hbLBD(1-84), a TEV cleavage site in the linker region, and hbSBD
(104-152), generously provided to us by Dr. D.T. Chuang of the
Southwestern Medical Center, University of Texas.  There is an
extra glycine in front of Glu104 which is left over after TEV
cleavage, and extra leucine,
 glutamic acid at the C-terminus before six histidine residues.
 The protein was purified by Ni-NTA affinity chromatography, and the
 purity of the protein was found to be better than 95\%,
based on the Coomassie blue-stained gel. The complete sequence of
$N=52$ residues for hbSBD is\\
(G)EIKGRKTLATPAVRRLAMENNIKLSEVVGSGKDGRILKEDILNYLEKQT(L)(E).

\vskip 2 mm

\noindent{\bf Circular Dichroism}

\vskip 2 mm

\noindent
 CD measurements were carried out in Aviv CD spectrometer model 202
with temperature and stir control units at different temperature
taken from
 260nm to 195nm. All experiments were carried at 1 nm bandwidth
 in 1.0 cm quartz square cuvette thermostated to  $\pm 0.1^o$C.
Protein concentration ($\sim$ 50 uM) was determined by UV absorbance at 280nm
using $\epsilon _{280nm}$=1280 M$^{-1}$cm$^{-1}$ with 50mM phosphate buffer at pH7.5.
 Temperature control was achieved using a circulating water bath system,
 and the equilibrium time was three minutes for each temperature point.
The data
was collected at each 2K increment in temperature. The study was
 performed at heating rate
of 10$^{o}$C/min and equilibration time of 3 minutes.
 The volume changes as a result of thermal expansion as well
 as evaporation of water were neglected.

\vskip 2 mm \noindent {\bf Fitting Procedure}

\vskip 2 mm

\noindent
 Suppose the thermal denaturation is a two-state
transition, we can write the ellipticity as
\begin{equation}
\theta \; \, = \; \, \theta _D + (\theta _N - \theta _D)f_N \, ,
\label{theta_fN_eq}
\end{equation}
where $\theta _D$ and $\theta _N$ are values for the denaturated
and folded states. The fraction of the folded conformation
$f_N$ is expressed as  \cite{Privalov79}
\begin{eqnarray}
f_N \; \, &=& \; \, \frac{1}{1 + \exp (-\Delta G_T/T)} \, ,\nonumber \\
\Delta G_T \; \, &=& \; \, \Delta H_T - T\Delta S_T \; = \;
\Delta H_G\left(1 - \frac{T}{T_G}\right) \nonumber \\
&+&\Delta C_p \left[(T-T_G) - T \ln \frac{T}{T_G}\right] \, .
\label{fN_twostate_eq}
\end{eqnarray}
Here $\Delta H_G$ and $\Delta C_p$ are jumps of the enthalpy
and heat capacity at the mid-point temperature $T_G$ (also known
as melting or
folding temperature) of thermal transition, respectively.
Some other thermodynamic characterization of stability
such as the temperature of maximum stability ($T_S$), the
temperature with zero enthalpy ($T_H$), and the conformational
stability ($\Delta G_S$) at $T_S$ can be computed
from results of regression analysis \cite{Becktel87}
\begin{eqnarray}
\ln \frac{T_G}{T_S} \; &=& \; \frac{\Delta H_G}{T_G\Delta C_p}, \\
T_H \; &=& \; T_G - \frac{\Delta H_G}{\Delta C_p}, \\
\Delta G_S \; &=& \; \Delta C_p (T_S - T_H).
\label{parameters_eq}
\end{eqnarray}
Using Eq. (\ref{theta_fN_eq}) - Eq. (\ref{parameters_eq})
we can obtain all thermodynamic parameters from CD data.

It should be noted that the fitting of Eq. \ref{fN_twostate_eq}
with $\Delta C_p > 0$ allows for an additional cold denaturation
\cite{Privalov90} at temperatures much lower than the room
temperature . The temperature of such a transition, $T_G'$, may be
obtained by the same fitting procedure with an additional
constraint of $\Delta H_G <0$. Since the cold denaturation
transition is not seen in Go models, to compare the simulation
results to the experimental ones we also use the approximation in
which $\Delta C_p=0$.

\vskip 2 mm \noindent {\bf Simulation} \vskip 2 mm

\noindent
We use coarse-grained continuum representation for hbSBD
protein, in which only the positions of 52 C$_{\alpha}$-carbons
are retained. We adopt the off-lattice version of the Go model
\cite{Go} where the interaction between residues forming native
contacts is assumed to be attractive and the non-native
interactions - repulsive. Thus, by definition for the Go model the
PDB structure is the native
 structure with the lowest energy.
The advantage of this model  is its simplicity which allows one to
study model proteins in detail.
Following Ref. \onlinecite{Clementi00}, we write
the energy of the Go-like model as
\begin{eqnarray}
&E& \; = \; \sum_{bonds} K_r (r_{i,i+1} - r_{0i,i+1})^2 + \sum_{angles} K_{\theta}
(\theta_i - \theta_{0i})^2 \nonumber \\
&+& \sum_{dihedral} \{ K_{\phi}^{(1)}
[1 - \cos (\Delta \phi_i)] +
K_{\phi}^{(3)} [1 - \cos 3(\Delta \phi_i)] \} \nonumber\\
&+& \sum_{i>j-3}^{NC}  \epsilon_1 \left[ 5 R_{ij}^{12} - 6 R_{ij}^{10}\right] +
\sum_{i>j-3}^{NNC} \epsilon_2 \left(\frac{C}{r_{ij}}\right)^{12} .
\label{Hamiltonian}
\end{eqnarray}
Here $\Delta \phi_i=\phi_i - \phi_{0i}$, $R_{ij}={r_{0ij}}/{r_{ij}}$;
$r_{i,i+1}$, $\theta_i$ and $\phi_i$ stand for the $i$th bond length between
 the $i$th and $(i+1)$th residues, the bond angle
 between the $(i-1)$th and $i$th bonds
and the dihedral angle around the $i$th bond, respectively.
$r_{ij}$ is the distance between the $i$th and $j$th residues.
Subscript ``0'', ``NC'' and ``NNC'' refer to the native
conformation, native contacts and non-native contacts,
respectively. The first harmonic term keeps the chain
connectivity, while the second and third terms represent the local
 angular interactions. Two last terms are non-local interactions, where
the former includes native contact interactions and the latter is
nonspecific repulsion between non-native pairs. We choose $K_r =
100 \epsilon_H$, $K_{\theta} = 20 \epsilon_H,
 K_{\phi}^{(1)} = \epsilon_H,
K_{\phi}^{(3)} = 0.5\epsilon_H, \epsilon_1 = \epsilon_H
, \epsilon_2 = \epsilon_H$ and $C = 4$ \AA, where $\epsilon_H$ is the
characteristic hydrogen bond energy \cite{Clementi00}.

The nativeness of any configuration is measured by the number of
native contacts $Q$. We define that the $i$th and $j$th residues
are in the native contact if $r_{0ij}$ is smaller than a cutoff
distance $d_c$ taken to be $d_c = 7.5$ \AA,
where $r_{0ij}$ is the distance between the $i$th and $j$th residues in
the native conformation. Using this definition and the native
conformation of Ref. \onlinecite{ccf3}, we found that the total
number of native contacts $Q_{total}$ is $62$. To study the
probability of being in the native state we use the following
overlap function \cite{Camacho93PNAS}
\begin{equation}
\chi \; = \frac{1}{Q_{total}} \sum_{i<j+1}^N \,\;
\theta (1.2r_{0ij} - r_{ij}) \Delta_{ij}
\label{chi_eq}
\end{equation}
where
$\Delta_{ij}$ is equal to 1 if residues $i$ and $j$ form a native
contact and 0 otherwise and $\theta (x)$ is the Heaviside
function. The argument of this function guarantees that
a native contact between $i$ and $j$ is classified as formed
when $r_{ij}$ is shorter than 1.2$r_{0ij}$ \cite{Clementi00}

The overlap function $\chi$, which is one if the
conformation of the polypeptide chain coincides with the native
structure and zero for unfolded conformations, can serve as an
order parameter for the folding-unfolding transition. The
probability of being in the native state, $f_N$, which can be
measured by the CD and other experimental techniques, is defined
as $f_N = <\chi>$, where $<...>$ stands for a thermal average.

The dynamics of the system is obtained by integrating the following Langevin
equation \cite{Allen_book}
\begin{equation}
m\frac{d^2\vec{r}}{dt^2} \; \; = \; \;
- \zeta \frac{d\vec{r}}{dt} + \vec{F}_c + \vec{\Gamma},
\label{DynaEq_eq}
\end{equation}
where $m$ is the mass of a bead, $\zeta$ is the friction coefficient,
$\vec{F}_c = dE/d\vec{r}$. The random force $\vec{\Gamma}$ is a white noise,
i.e. $<\Gamma_i(t) \Gamma_j(t')> = 2\zeta k_BT\delta_{ij}\delta(t-t')$,
where $i$ and $j$ refer to components $x,y$ and $z$.
It should be noted that the folding thermodynamics
does not depend on the enviroment viscosity (or on $\zeta$)
but the folding kinetics depends
on it \cite{KlimThirumPRL97}.  We chose the dimensionless parameter
$\tilde{\zeta} = (\frac{a^2}{m\epsilon_H})^{1/2}\zeta = 8$, where
$m$ is the mass of a bead and $a$ is the bond length between successive beads.
One can show that this value of $\tilde{\zeta}$ belongs to the
interval of the viscosity where the folding
kinetics is fast. We have tried other values of $\tilde{\zeta}$
but the results
remain unchanged qualitatively.

We measure time in units of $\tau_L = (ma^2/\epsilon_H)^{1/2}$.
Using the typical value $m = 3.10^{-25}$ kg \cite{Veitshans97},
$a = 4$\AA $~$ and $\epsilon_H = 0.91$ kcal/mol 
(this choice of $\epsilon_H$ follows from the requirement
that the simulated folding temperature coincides
with the experimental one, see below)
we obtained $\tau_L \approx 3$ ps.
The dynamics Eq. (\ref{DynaEq_eq}) was solved by the Verlet
algorithm \cite{Swope82} with the time step $\Delta t = 0.005 \tau_L$.
All thermodynamic quantities are obtained by the histogram method
\cite{Ferrenberg89}.


\vskip 6 mm
\noindent{\Large \bf Results}
\vskip 2 mm

\noindent{\bf CD  Experiments} \vskip 2 mm

\noindent
 The structure of hbSBD
is shown in Figure 1. Its conformational stability is investigated
in present study by analyzing the unfolding transition induced by
temperature as monitored
 by CD, similar to that described previously \cite{Naik02,Naik04}.
The reversibility of thermal denaturation was ascertained by monitoring
the return of the CD signal upon cooling from 95$^{o}$C to 22 $^{o}$C;
immediately after the conclusion of the thermal transition.
The transition was found to be more than 80\% reversible.
 Loss in reversibility to greater extent was observed on prolonged
 exposure of the sample to higher temperatures.
 This loss of reversibility is presumably due to irreversible
 aggregation or decomposition. Figure 2
 shows the wavelength dependence
 of mean residue molar
ellipticity  of hbSBD at various temperatures between 278K and 363K.
 In a separate study, the thermal unfolding transition as monitored
 by ellipticity at 228 nm was found to be independent of hbSBD
 concentration in the range of 2 uM to 10 uM. It was also found to be
 unaffected by change in heating rate between 2$^{o}$C/min to 20$^{o}$C/min.
These observations suggest absence of stable intermediates in heat
 induced denaturation of hbSBD. A valley at around 220 nm,
 characteristics of the helical secondary structure is evident for
 hbSBD.


Figure 3 shows the temperature dependence of the
population of the native conformation, $f_N$, for wave lengths
$\lambda = 208, 212$ and 222 nm. We first try to fit these data to
Eq. (\ref{fN_twostate_eq}) with $\Delta C_p \ne 0$. The fitting
procedure gives slightly different values for the folding (or
melting) temperature and the enthalpy jump for three sets of
parameters. Averaging over three values, we obtain $ T_G = 317.8
\pm 1.95$ K and $\Delta H_G = 19.67 \pm 2.67$ kcal/mol. Other
thermodynamic quantities are shown on the first row of Table 1.
The similar fit but with $\Delta C_p=0$ gives the
thermodynamic parameters shown on the second row of this table.
Since the experimental data are nicely fitted to the two-state
model we expect that the downhill scenario does not applied to the
hbSBD domain.


For the experimentally studied temperature interval two types of
the two-state fit (\ref{fN_twostate_eq}) with $\Delta C_p=0$ and
$\Delta C_p \ne 0$ give almost the same values for $T_G$, $\Delta
H_G$ and $\Delta S_G$. However, pronounced different behaviors of
the population of the native basin, $f_N$, occur when we
interpolate results to the low temperature region (Fig. 4).
For the $\Delta C_p=0$ case, $f_N$ approaches
the unity as $T \rightarrow 0$ but it goes down for $\Delta C_p
\ne 0$. This means that the $\Delta C_p \ne 0$ fit is valid if the
second cold denaturation transition may occur at $T_G$'. This
phenomenon was observed in single domains as well as in
multi-domain globular proteins \cite{Privalov90}. We predict that
the cold denaturation of hbSBD may take place at $T_G' \approx
212$ K which is lower than $T_G' \approx 249.8$ K for hbLBD
shown on the 4th row of Table 1.
It would be of great interest to carry out the cold denaturation
experiments in cryo-solvent to elucidate this issue.

To compare the stability of the hbSBD domain with the hbLBD domain
which has been studied in detail previously \cite{Naik04} we also
present the thermodynamic data of the latter on Table 1. Clearly,
hbSBD is less stable than hbLBD by its smaller $\Delta G_S$ and
lower $T_G$ values. This is consistent with their respective
backbone dynamics as revealed by $^{15}$N-T$_1$, $^{15}$N-T$_2$,
and  $^{15}$N-$^1$H NOE studies of these two domains using
uniformly  $^{15}$N-labeled protein samples (Chang and Huang,
unpublished results). Biologically, hbSBD must bind to either E1
or E3 at different stages of the catalytic cycle, thus it needs to
be flexible to adapt to local enviroments of the active sites of
E1 and E3. On the other hand, the function of hbLBD is to permit
its Lys44 residue to channel acetyl group between donor and
acceptor molecules and only the Lys44 residue needs to be flexible
\cite{Chang_JBC02}. In addition, the NMR observation for the
longer fragment (comprising residues 1-168 of the E2 component)
also showed that the hbLBD region would remain structured after
several months while the hbSBD domain could de-grate in a shorter
time.



\vskip 2 mm \noindent {\bf Folding Thermodynamics from
Simulations}

\vskip 2 mm

\noindent

In order to calculate the thermodynamics quantities we have collected
histograms for the energy and native contacts
at six values of temperature: $T = 0.4, 0.5, 0.6, 0.7,0.8$
and 1.0 $\epsilon_H/k_B$. For sampling,
at each temperature 30 trajectories
of $16\times 10^7$ time steps have been generated with initial
$4\times 10^7$ steps discarded for thermalization. 
The reweighting histogram method \cite{Ferrenberg89} was used 
to obtain the thermodynamics parameters at all temperatures.

 Figure 4 (open circles) shows the temperature
dependence of population of the native state, defined as the
renormalized number of native contacts (see Material and Methods)
for the Go model. Since there is no cold denaturation for this model,
to obtain the thermodynamic parameters we fit $f_N$ to the
two-state model (Eq. \ref{fN_twostate_eq}) with $\Delta C_p=0$.

The fit (black curve) works pretty well around the transition
temperature but it gets worse at high $T$ due to slow decay of
$f_N$ which is characteristic for almost all of theoretical
models. In fitting we have chosen the hydrogen bond energy
$\epsilon_H = 0.91$ kcal/mol in Hamiltonian (\ref{Hamiltonian}) so
that $T_G = 0.7 \epsilon_H/k_B$ coincides with
 the experimental value 317.8 K. From the
fit we obtain $\Delta H_G = 11.46$ kcal/mol which is smaller than
the experimental value indicating that the Go model is less
stable compared to the real hbSBD.

Figure 5 shows the temperature dependence of derivative of the
fraction of native contacts with respect to temperature  $df_N/dT$
(we also call this value the structural susceptibility) and the
specific heat $C_v$ obtained from the Go simulations. The collapse
temperature $T_{\theta}$, defined as the temperature at which
$C_v$ is maximal, almost coincides with the folding temperature
$T_F$ (at $T_F$ the structural susceptibility has maximum).
According to Klimov and Thirumalai \cite{KlimThirum96PRL},
the dimensionless parameter $\sigma = \frac{|T_{\theta}-T_F|}{T_F}$
may serve as an indicator for foldablity of proteins. Namely,
sequences with $\sigma \leq 0.1$ fold much faster that  
those which have the same number of residues but with $\sigma$
exceeding 0.5. From this perspective, having $\sigma \approx 0$ hbSBD
is supposed to be a
good folder {\it in silico}. However, one has to be cautious about
this conclusion because 
the pronounced correlation between folding times $\tau _F$
and the equilibrium parameter
$\sigma$, observed for simple on- and off-lattice models 
\cite{KlimThirum96PRL,Veitshans97} may be not valid for proteins in laboratory
\cite{Plaxco_Rev04}. In our opinion, since the data collected from
theoretical and
experimental studies are limited, further studies are required to
clarify the relationship between $\tau _F$ and $\sigma$.    

 Using experimental values for
$T_G$ (as $T_F$) and $\Delta H_G$ and the two-state model with $\Delta C_p
=0$ (see Table 1) we can obtain the temperature dependence of the
population of native state $f_N$ and, therefore, $df_N/dT$ for
hbSBD (Fig. 5). Clearly, the folding-unfolding transition
{\it in vitro} is sharper than
in the Go modeling. One of possible reasons is that our Go
model ignores the side chain which can enhance the cooperativity of
the denaturation transition \cite{KlimThirum98FD}.


The sharpness of the fold-unfolded transition might be characterized
quantitatively
via the cooperativity index $\Omega _c$ which is defined as follows
\cite{Li04}
\begin{equation}
 \Omega_c=\frac{T_F^2}{\Delta T}
\biggl(\frac{df_N}{dT}\biggr)_{T=T_F},
\end{equation}
where $\Delta T$ is the transition width. From Fig. 5, we obtain
$\Omega_c = 51.6$ and 71.3 for the Go model and CD experiments,
respectively. Given the simplicity of the Go model used here the
agreement in $\Omega_c$ should be considered reasonable. We can
also estimate $\Omega _c$ from the scaling law suggested in Ref.
\onlinecite{Li04a}, $\Omega _c = 0.0057 \times N^{\mu}$, where
exponent $\mu$ is universal and expressed via the random walk
susceptibility exponent $\gamma$ as $\mu = 1+\gamma
\approx 2.22 (\gamma
\approx 1.22$). Then we get $\Omega_c \approx 36.7$ which is lower
than the experimental as well as simulation result. This means
that hbSBD {\em in vitro} is, on average, more cooperative than
other two-state folders.

Another measure for the cooperativity is
$\kappa _2$ which is defined as  \cite{Chan00PRL}
$\kappa _2 = \Delta H_{vh}/\Delta H_{cal}$, where
$\Delta H_{vh} \;  = \; 2T_{max}\sqrt{k_B C_V(T_{max})}$
and $\Delta H_{cal} \; = \;  \int_0^{\infty} C_V(T)dT$,
are the van't Hoff and the calorimetric enthalpy, respectively,
$C_V(T)$ is the specific heat. Without the baseline substraction
in $C_V(T)$ \cite{Chan_ME04}, for the Go model of hbSBD we
obtained $\kappa _2 \approx 0.25$. Applying the baseline
substraction
as shown in the lower part of Fig. 5
we got $\kappa _2 \approx 0.5$ which is still much lower than
$\kappa _2 \approx 1$ for a trully all-or-none transition.
Since $\kappa _2$ is an extensive parameter, its low value
is due to the shortcomings of the off-lattice
Go models but not due to the finite size effects.
More rigid lattice models give better results for
the calorimetric cooperativity \cite{Li_Physica05}.
Thus, for the hbSBD domain the Go model gives
the better agreement with our CD experiments for the
structural cooperativity
$\Omega_c$ than for the calorimetric measure $\kappa _2$.

\vskip 2 mm
\noindent{\bf Free Energy Profile}
\vskip 2 mm

\noindent
To get more evidence that hbSBD is a two-state folder we
study the free energy profile using some quantity as a reaction
coordinate. The precise reaction coordinate for a
multi-dimensional process such as protein folding is difficult to
ascertain. However, Onuchic and coworkers \cite{Nymeyer98PNAS}
have argued that, for minimally frustrated systems such as Go
models, the number of native contact $Q$ may be appropriate. Fig.
6(a) shows the dependence of free energy on $Q$ for $T=T_F$. Since
there is only one local maximum corresponding to the transition
state (TS), hbSBD is a two-state folder. This is not unexpectable
for hbSBD which contains only helices. The fact that the simple Go model
correctly captures the two-state behavior as was observed in the CD
experiments, suggests that the energetic frustration ignored in this model
plays a minor role compared to the topological frustration
\cite{Clementi00}.


We have sorted out structures of the denaturated state (DS), TS
and the folded state (FS) at $T=T_F$ generating $10^4$
conformations in equilibrium. The distributions of the RMSD,
$P_{\rm RMSD}$,
 of these states are
plotted in  Fig. 6(b). As expected, $P_{\rm RMSD}$ for
the DS spreads out more than that for the TS and FS. According to
the free energy profile in Fig. 6(a), the TS conformations
have 26 - 40 native contacts. We have found that the size (number
of folded residues) \cite{Bai04} of the TS is equal to 32. Comparing this size
with the total number of residues ($N=52$) we see that the fraction of
folded residues in the TS is higher than the typical value
for real two-state proteins
\cite{Bai04}. This is probably an artefact of Go models \cite{LiKouza05}.
The TS conformations are relatively compact having
the ratio $<R_g^{TS}>/R_g^{NS} \approx 1.14$, where $<R_g^{TS}>$
is the average radius of gyration of the TS ensemble and $R_g^{NS}$
is the radius of gyration of the native conformation shown in Fig. 1.
Since the RMSD, calculated only for two helices, is about 0.8 \AA
the structures of two helices in the TS
are not distorted much. It is also evident from
the typical structure of the TS shown in Fig.6(b) where
the helix regions H$_1$ and H$_2$ involve residues 13 - 19 and 39 - 48,
respectively (a residue is considered to be in the helix state if its
dihedral angle is about 60$^o$).
Note that H$_1$ has two residues less compared to
H$_1$ in the native conformation (see the caption to Fig. 1)
but H$_2$ has even one bead more than its native state counterpart.
Overall, the averaged RMSD of the TS conformations from the
native conformation (Fig. 1) is about
4.9 \AA$~$ indicating that the TS is not close to the native one. As seen
from Figs. 6(a) and 1, the main difference comes
from the tail parts. The most probable conformations
(corresponding to maximum of $P_{\rm RMSD}$ in Fig. 6(b) of the FS
have RMSD about 2.5 \AA. This value is reasonable from the point
of view of the experimental structure resolution.

\vskip 2 mm
\noindent {\bf Folding Kinetics}
\vskip 2 mm

\noindent
The two-state foldability, obtained from the thermodynamics simulations
may be also  probed by studying
the folding kinetics. For this purpose we monitored the
time dependence of the fraction of unfolded trajectories $P_u(t)$ defined
as follows \cite{Klimov99}
\begin{equation}
P_u(t) \, = \, 1 - \int_0^t P^{\textstyle{N}}_{fp}(s)ds,
\label{Pu_eq}
\end{equation}
where $P^{\textstyle{N}}_{fp}$ is the distribution of first passage folding
times
\begin{equation}
P^{\textstyle{N}}_{fp} \, = \, \frac{1}{M} \sum_{i=1}^{M}
\delta (s - \tau _{f,1i}).
\label{Pn_eq}
\end{equation}
Here $\tau _{f,1i}$ is time for the $i$th trajectory
to reach the native state for the first time,
$M$ is the total number of trajectories used in simulations.
A trajectory is said to be folded if all of native contacts form. As seen from
Eqs \ref{Pu_eq} and \ref{Pn_eq},
$P_u(t)$ is the fraction of trajectories which do not reach
the native state at time $t$.
In the two-state scenario the folding becomes triggered after 
overcoming only one free energy barrier between the transition state
and the denaturated one. Therefore, 
$P_u(t)$ should be a single exponential,
i.e. $P_u(t) \sim \exp(-t/\tau_F)$ (a multi-exponential
behavior occurs in the case when the folding proceeds via intermediates)
\cite{Klimov99}. Since the function $P_u(t)$ can be measured
directly by a number of experimental techniques \cite{Greene04,Dyson05},
the single exponential kinetics of two-state folders
is supported by a large body of experimental work (see, i.e. Ref.
\cite{Naik02} and references there).

Fig. 7 shows the semi-logarithmic plot for $P_u(t)$ at $T=T_F$ for
the Go model. 
Since the single exponential fit works pretty well, one can expect
that intermediates do not occur on the folding pathways.
Thus, together with the thermodynamics data our kinetic study supports 
the two-state behavior of the hbSBD domain as observed
on the CD experiments.

From the linear fit in Fig. 7 we obtain the folding time
$\tau_F \approx 0.1 \mu$s.
This value is consistent with the
estimate of the folding time defined as the average value of the
first passage times.
If we use the empirical formula for the folding time
$\tau_F = \tau_F^0 \exp(1.1N^{1/2})$, where prefactor
$\tau_F^0=0.4 \mu$s and $N$ is a number of amino acids \cite{Li04}
then $\tau_F = 1.1\times 10^3 \mu$s for $N=52$. This value is
about four orders of magnitude larger than that obtained from the
Go model.
 Thus the Go model can capture the two-state feature of
the denaturation transition for hbSBD domain but not folding
times.


\vskip 6 mm
\noindent{\Large \bf Discusion}
\vskip 2 mm

We have used CD technique and the Langevin dynamics to study the
mechanism of folding of hbSBD. Our results suggest that this
domain is a two-state folder. The CD experiments reveal that the
hbSBD domain is less  stable than the hbLBD domain in the same
BCKD complex, but it is more stable and cooperative compared to
other fast folding $\alpha$ proteins. 

Both the thermodynamics and
kinetics results, obtained from the Langevin dynamics
simulations, show that the simple Go model correctly captures
the two-state feature of folding.
It should be noted that the two-state behavior is not the natural
consequence of the Go modeling because it allows for fishing folding
intermediates caused by the topological frustration. From this standpoint
it may be used to decipher the foldability of model proteins
for which the topological frustration dominates.
The reasonable agreement between
the results obtained by the Go modeling 
and our CD experiments,
suggests that the native state topology of hbSBD is more important
than the energetic factor. 

The theoretical model gives the reasonable
agreement with the CD experimental data for the structural
cooperativity $\Omega _c$. However, the calorimetric cooperativity
criterion $\kappa _2 \approx 1$ for two-state folders is hard to
fulfill within the Go model. From the $\Delta C_p \ne 0$ fitting
procedure we predict that the cold denaturation of hbSBD may occur
at $T \approx 212$ K and it would be very interesting to verify
this prediction experimentally. We are using the package SMMP
\cite{smmp,smmpnew} and a parallel algorithm \cite{jcc01} to
perform all-atom simulation of hbSBD to check the relevant
results. \vskip 2 mm

This work was
supported by KBN grant number 1P03B01827 and
 by National Science Council in Taiwan under grant numbers
No. NSC 93-2112-M-001-027 (to CKH) and No. NSC 92-2113-M-001-056
(to THH).
The NMR spectra used to determine hbSBD structure
were obtained at the High-field NMR Core Facility,
National Research Program for Genomic Medicine (NRPGM),
Taiwan, Republic of China.

\vspace{0.2cm}


\newpage
\vskip 20 mm

\vskip 10 mm

\begin{figure}
\epsfxsize=6.5in
\vspace{0.3in}
\centerline{\epsffile{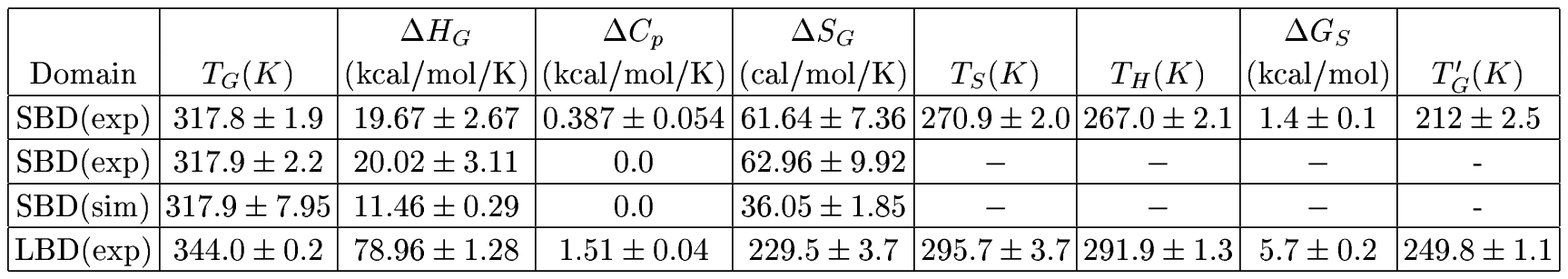}} \vspace{0.2in}
\end{figure}
\noindent {\bf TABLE 1.} Thermodynamic parameters obtained from
the CD experiments and simulations for hbSBD domain. The results
shown on the first and fourth rows were obtained by fitting
experimental data to the two-state equation (\ref{fN_twostate_eq})
with $\Delta C_p \ne 0$. The second and third rows corresponding
to the fit with $\Delta C_p = 0$. The results for hbLBD are taken
from Ref. \onlinecite{Naik04} for comparison.



\vskip 2 mm

\newpage

\centerline{\Large \bf Figure Captions} \vskip 5 mm

\noindent {\bf FIGURE 1.} Ribbon representation of the structure
of hbSBD domain. The helix region H$_{1}$ and H$_2$ include residues
Pro12 - Glu20 and Lys39 - Glu47, respectively.  \vskip 5 mm

\noindent {\bf FIGURE 2.} Dependence of the mean residue molar
ellipticity on the wave length for 18 values of temperatures
between 278 and 363 K. \vskip 5 mm

\noindent {\bf FIGURE 3.} Temperature dependence of the fraction
of folded conformations $f_N$, obtained from the ellipticity
$\theta$ by Eq. (\ref{fN_twostate_eq}), for wave lengths $\lambda$
= 208 (blue circles), 212 (red squares)  and 222 nm (green
diamonds). The solid lines corresponds to the two state fit given
by Eq. \ref{fN_twostate_eq} with $\Delta C_p \ne 0$. We obtained
$T_G=T_F= 317.8 \pm 1.9$ K, $\Delta H_G = 19.67 \pm 2.67$ kcal/mol
and $\Delta C_p = 0.387\pm0.054$.\vskip 5 mm

\noindent {\bf FIGURE 4.} The dependence of $f_N$ for various sets
of parameters. The blue and red curves correspond to the
thermodynamic parameters presented on the first and the second
rows of Table 1, respectively. Open circles refer to simulation
results for the Go model. The solid black curve is the two-state
fit ($\Delta C_p=0$) which gives $\Delta H_G= 11.46$ kcal/mol and
$T_F=317.9$.\vskip 5 mm

\noindent
 {\bf FIGURE 5.} The upper part refers to the temperature
dependence of $df_N/dT$ obtained by the simulations (red) and the
CD experiments (blue). The experimental curve is plotted using
two-state parameters with $\Delta C_p = 0$ (see, the second row on
Table 1). The temperature dependence of the heat capacity $C_V(T)$
is presented in the lower part. The dotted lines illustrate the
base line substraction. The results are averaged over 20 samples.
\vskip 5 mm

\noindent {\bf FIGURE 6.} (a) The dependence of free energy on the
number of native contacts $Q$ at $T=T_F$. The typical structures
of the denaturated state , transition state and folded
state are also drawn. The helix regions
H$_1$ (green) and H$_2$ (orange) of the TS structure involve residues 13 - 19
and 39 - 48, respectively. For the FS structure H$_2$ is the same as for
the TS structure but H$_1$ has two residues more (13 - 21). 
(b) Distributions of RMSD  for three
ensembles shown in (a). The average values of RMSD are equal to 9.8,
4.9 and 3.2 \AA$~$ for the DS, TS and FS, respectively.

\vskip 5 mm

\noindent {\bf FIGURE 7.} The semi-logarithmic plot of the time
dependence of the fraction of unfolded trajectories at $T=T_F$.
The distribution $P_u(t)$
was obtained from first passage times of 400 trajectories, which
start from random conformations.
The straight line corresponds to the fit ln $P_{\rm u}(t) =
-t/\tau_F$, where $\tau _F = 0.1 \mu$s.

\newpage

\begin{figure}
\epsfxsize=4.5in \epsfxsize=3.8 in
\centerline{\epsffile{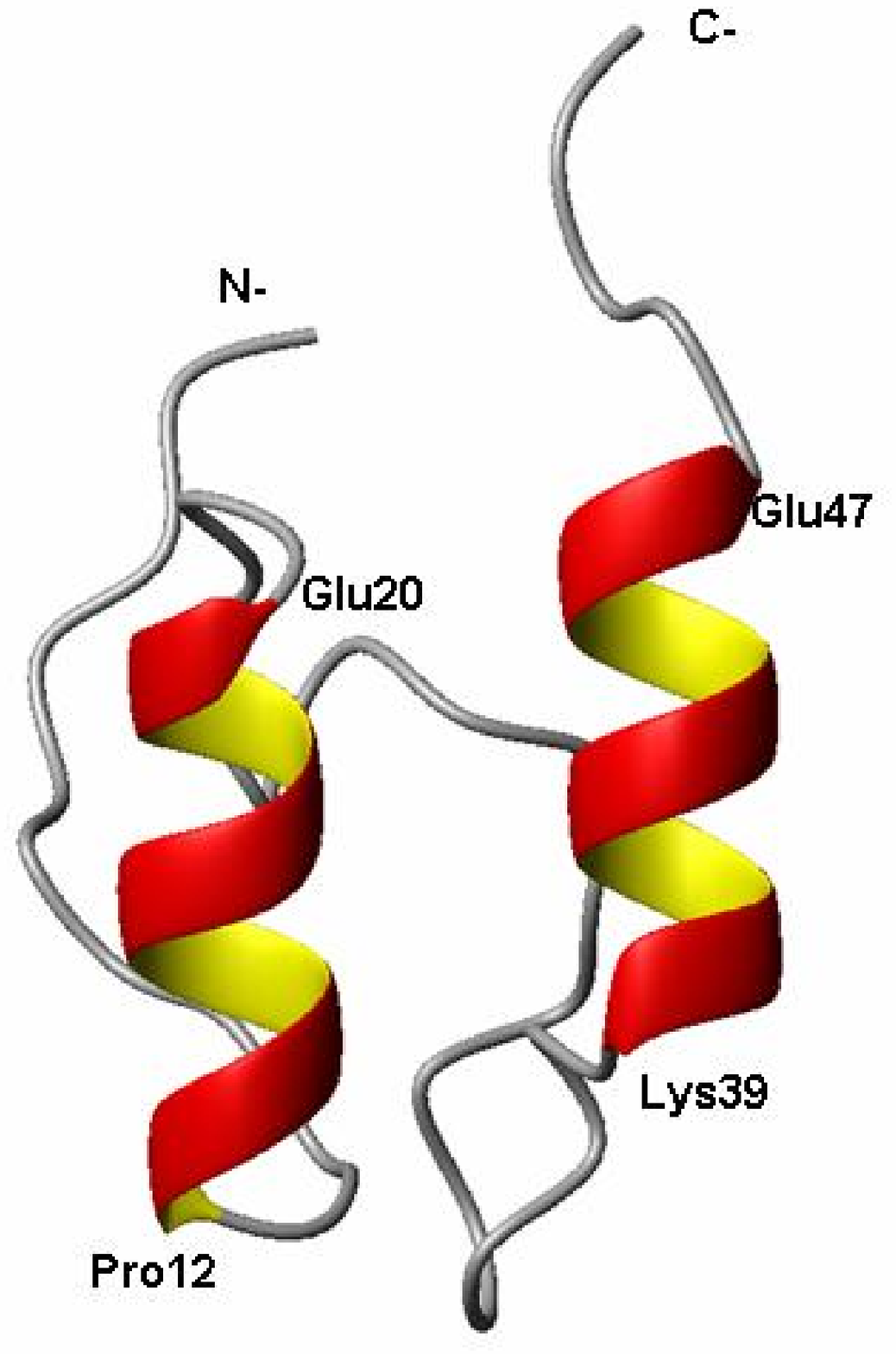}}
\centerline{\bf FIGURE 1.}
\end{figure}

\vskip 2 mm
\begin{figure}
\epsfxsize=3.5in \vspace{0.6in}
\centerline{\epsffile{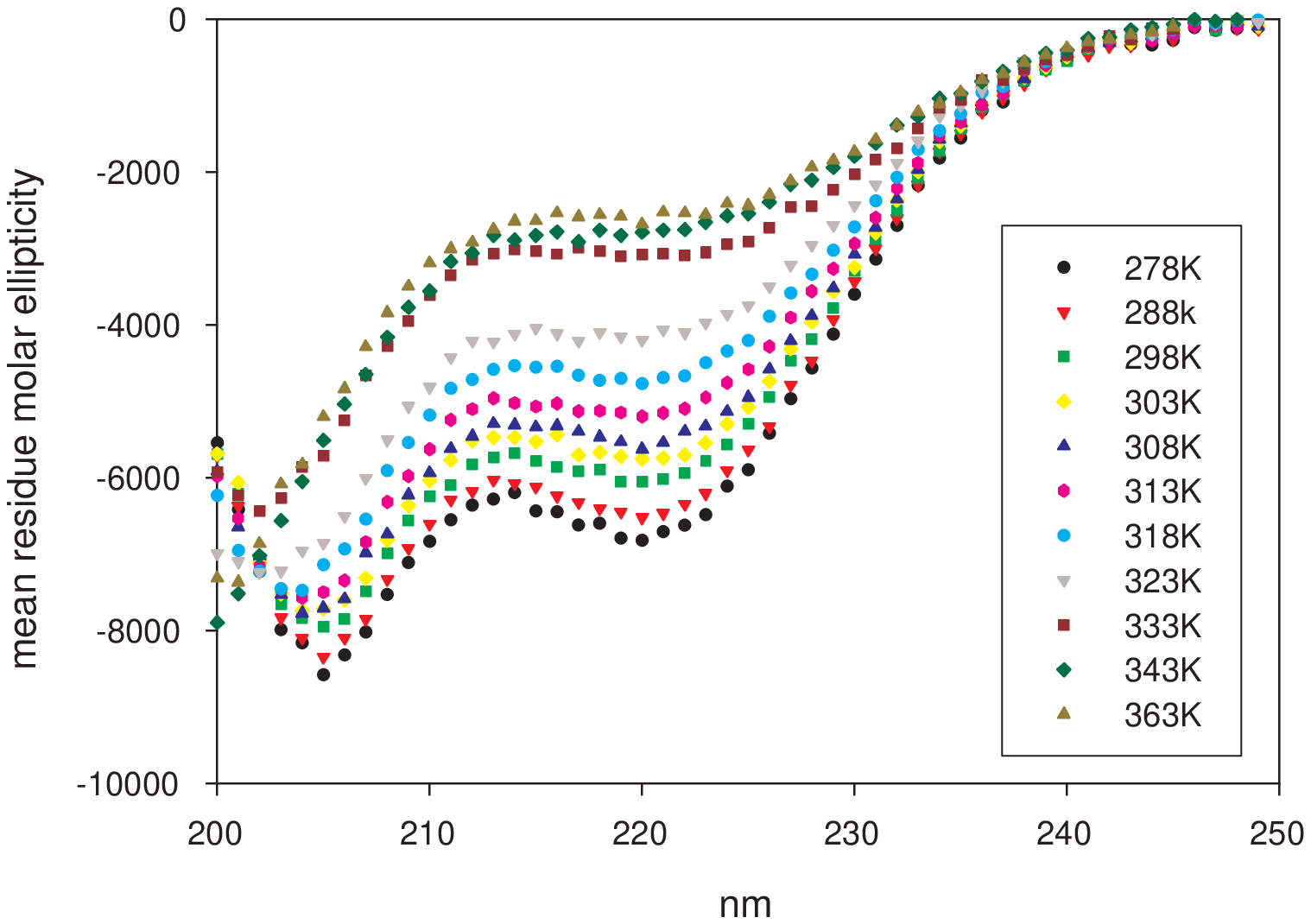}} \vspace{0.2in}
\end{figure}
\centerline{\bf FIGURE 2.}

\newpage
\vskip 2 mm
\begin{figure}
\epsfxsize=3.5in \vspace{0.3in}
\centerline{\epsffile{hbSBDfig3.eps}} \vspace{0.4in}
\end{figure}
\centerline{\bf FIGURE 3.}
\newpage
\vskip 2 mm
\begin{figure}
\epsfxsize=3.2in \vspace{0.3in}
\centerline{\epsffile{hbSBDfig4.eps}} \vspace{0.2in}
\end{figure}
\centerline{\bf FIGURE 4.}
\newpage
\vskip 2 mm
\begin{figure}
\epsfxsize=3.5in \vspace{0.3in}
\centerline{\epsffile{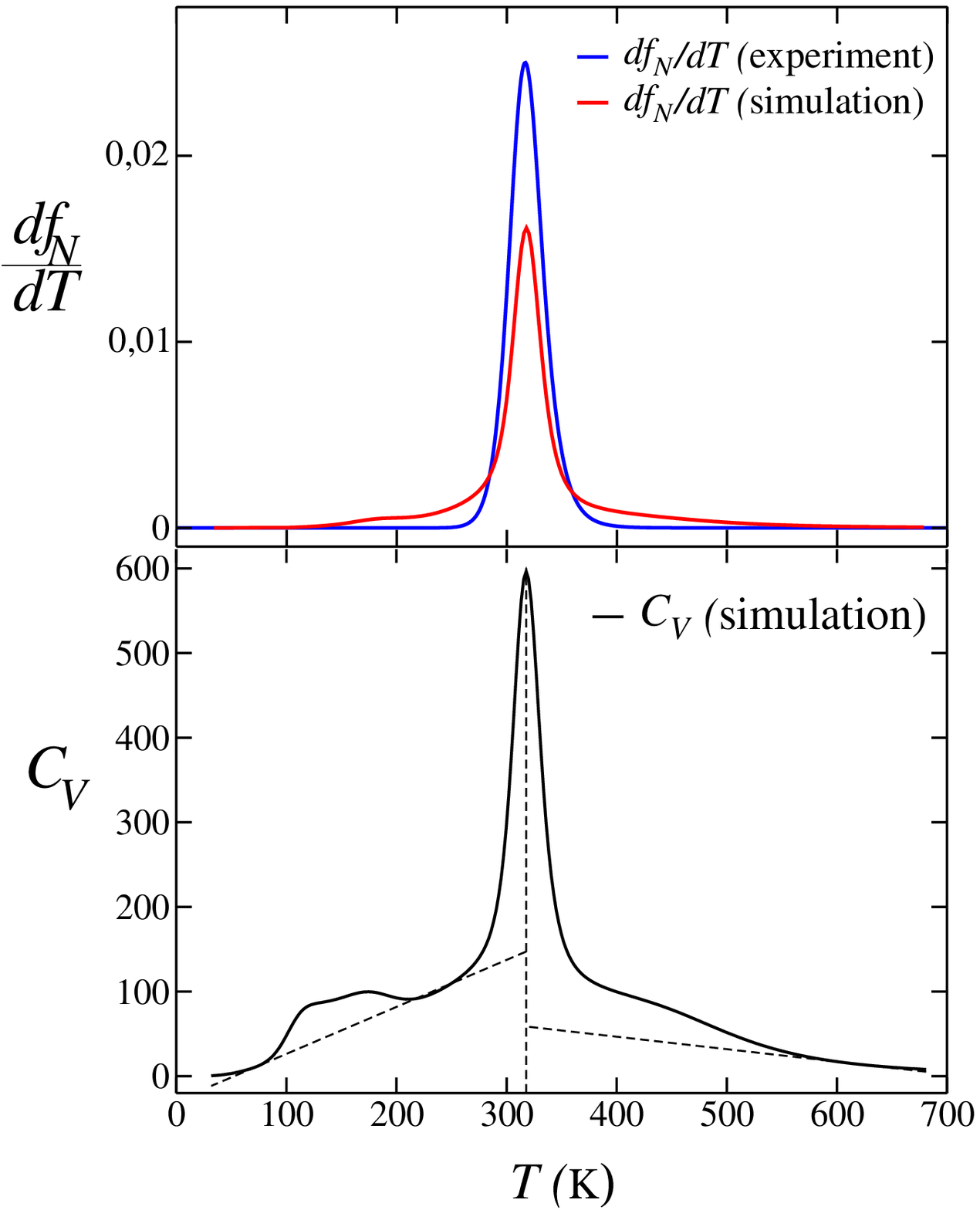}} \vspace{0.2in}
\end{figure}
\centerline{\bf FIGURE 5.}
\vskip 2 mm

\newpage

\begin{figure}
\epsfxsize=6.2in \epsfxsize=5.0 in \vspace{0.3in}
\centerline{\epsffile{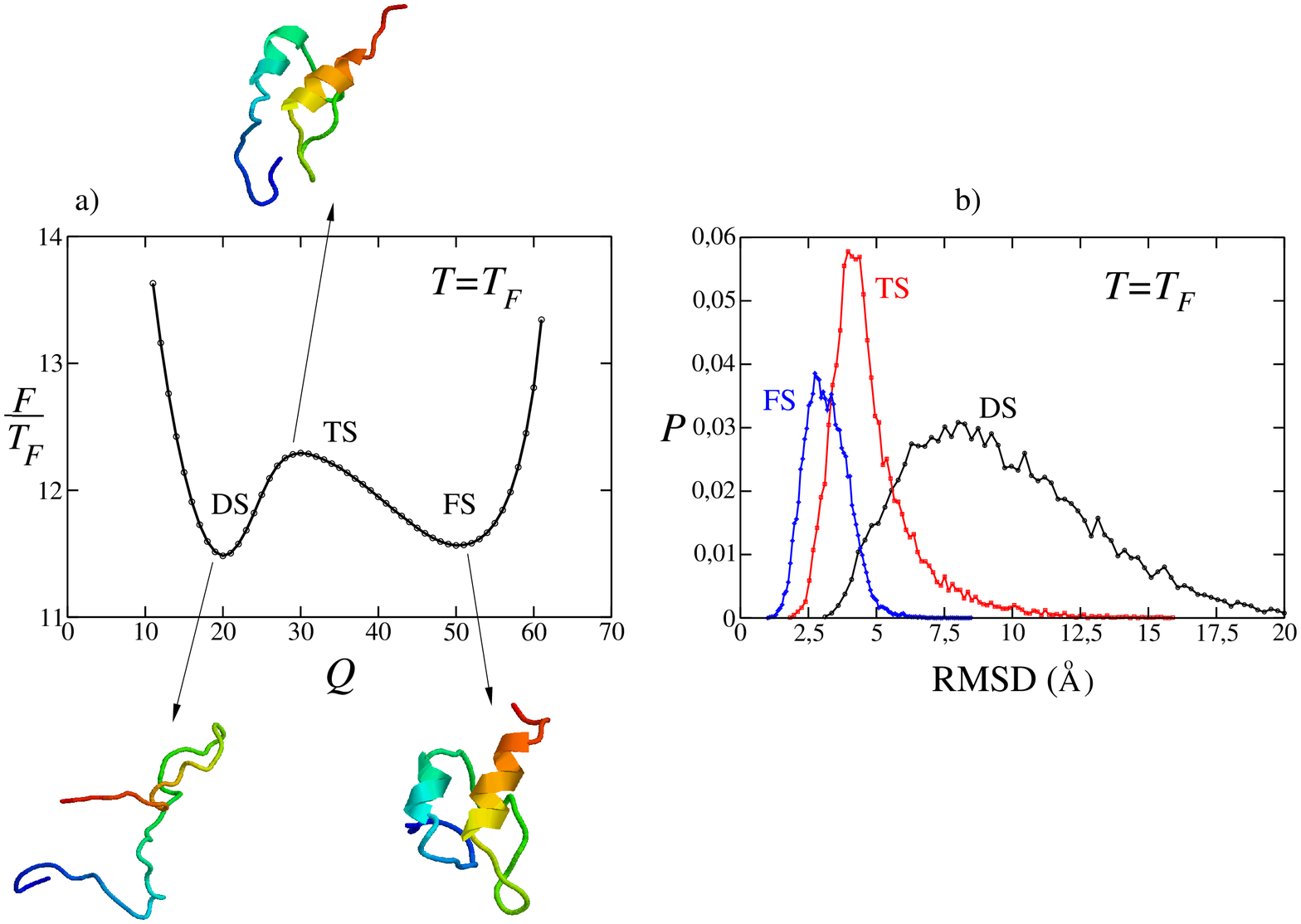}} \vspace{0.2in}
\label{free_fig}
\end{figure}
\centerline{\bf FIGURE 6.}

\newpage

\vskip 2 mm
\begin{figure}
\epsfxsize=3.5in \vspace{0.3in}
\centerline{\epsffile{hbSBDfig7.eps}} \vspace{0.2in}
\end{figure}
\centerline{\bf FIGURE 7.}



\end{document}